\documentclass[letterpaper]{article} 
\usepackage{aaai2027}  
\usepackage[hyphens]{url}  
\usepackage{graphicx} 
\usepackage{natbib}  
\usepackage{caption} 
\usepackage{algorithm}
\usepackage{algorithmic}

\usepackage{newfloat}
\usepackage{listings}
\DeclareCaptionStyle{ruled}{labelfont=normalfont,labelsep=colon,strut=off} 
\floatstyle{ruled}
\newfloat{listing}{tb}{lst}{}
\floatname{listing}{Listing}

\usepackage{booktabs}

\usepackage{amsmath}
\usepackage{amssymb}
\usepackage{booktabs}
\usepackage{multirow}
\usepackage{xcolor}
\usepackage{graphicx}
\usepackage{listings}
\usepackage{url}
\usepackage{comment}
\usepackage{cite}
\usepackage{cleveref}

\usepackage[table]{xcolor}
\definecolor{circuitprover}{HTML}{CDE09A}

\newcommand{\toolname}{CircuitProver}
\newcommand{\runtimemark}{\textsuperscript{\(\lozenge\)}}

\title{Lean4HW: Agentic AI Theorem Proving with \\ Reusable Proof Libraries for Hardware Verification}
\title{Lean4HW: Agentic \underline{Lean} \underline{4} Theorem Proving with Proof \\ Knowledge Reuse for Parameterized \underline{H}ard\underline{w}are Verification}
\title{CircuitProver: Agentic Lean~4 Theorem Proving with Reusable \\ Circuit Proof Library for Hardware Verification}
\author{
    Written by AAAI Press Staff\textsuperscript{\rm 1}\thanks{With help from the AAAI Publications Committee.}\\
    AAAI Style Contributions by Peter Patel Schneider,
    Sunil Issar,\\
    J. Scott Penberthy,
    George Ferguson,
    Hans Guesgen,
    Francisco Cruz\equalcontrib\corresponding,
    Marc Pujol-Gonzalez\equalcontrib\corresponding
}
\affiliations{
    \textsuperscript{\rm 1}Association for the Advancement of Artificial Intelligence\\

    1101 Pennsylvania Ave, NW Suite 300\\
    Washington, DC 20004 USA\\
    proceedings-questions@aaai.org
}

\title{CircuitProver: Agentic Lean~4 Theorem Proving with Reusable \\ Circuit Proof Library for Hardware Verification}
\author {
    Ziyi Yang\textsuperscript{\rm 1}\equalcontrib,
    Wenji Fang\textsuperscript{\rm 2}\equalcontrib,
    Chen Chen\textsuperscript{\rm 1},
    Zhiyao Xie\textsuperscript{\rm 2},
    Hongce Zhang\textsuperscript{\rm 1,2}\corresponding
}
\affiliations {
    \textsuperscript{\rm 1}Hong Kong University of Science and Technology(Guangzhou)\\
    \textsuperscript{\rm 2}Hong Kong University of Science and Technology\\
}

\begin{document}

\maketitle

\begin{abstract}


Modern integrated circuits (ICs) are becoming increasingly complex,
making functional verification a major bottleneck. 
The dominant hardware formal verification methodology, model checking, verifies each design instance separately and exposes only pass/fail results, so the reasoning behind a proof stays locked inside solver heuristics and is repeatedly reconstructed across related designs. 
Interactive theorem proving instead yields explicit, reusable proof artifacts, but applying it to hardware remains largely manual, demanding expert effort for formalization, invariant discovery, and proof development.

In this paper, we present \textbf{\toolname{}}, an agentic Lean~4-based verification framework supporting proof-accumulation and parameterized verification. 
\toolname{} automatically translates parameterized hardware designs and their natural language specifications into executable Lean~4 models. 
It then iteratively constructs machine-checked proofs through Lean feedback to establish that the hardware code complies with the  specification.
The proving traces and verified theorems are distilled into reusable libraries, where proving strategies guide future agent reasoning and verified lemmas support formal proof reuse across related hardware verification tasks.

We further introduce the first benchmark suite for evaluating agentic hardware theorem proving, covering diverse parameterized hardware designs, specifications, proof tasks, and evaluation metrics. 
Across 63 tasks, \toolname{} successfully proves all benchmarks, while a vanilla agent solves 92.1\% of them and requires twice as many proof rounds on average. 
Ablation studies show that accumulated proof knowledge reduces redundant proof construction across related verification tasks, reducing proof length by 16.3\% and verification time by 23.2\%.
These results show that reusable proof knowledge can significantly reduce redundant proof efforts on related verification tasks.


\end{abstract}
\section{Introduction}


The rapid growth of artificial intelligence is creating unprecedented demand for new integrated circuits (ICs). To meet this demand, hardware designers need to develop increasingly complex designs within ever shorter development cycles. Modern IC design therefore increasingly relies on reusable IPs and AI-assisted methodologies, with autonomous design agents emerging as a promising paradigm for improving productivity. 
As autonomous agents begin to participate in hardware generation~\cite{yu2026agentic}, optimization~\cite{fang2026dr}, and debugging~\cite{bai2025fvdebug}, automated verification becomes increasingly important for validating their outputs and providing reliable feedback. 
However, ensuring functional correctness remains a major bottleneck~\cite{11539347}. 
While simulation can detect many bugs, it cannot exhaustively cover all possible execution behaviors, making formal verification indispensable for modern hardware development.\looseness=-1


\begin{figure}[!t]
  \centering
  \includegraphics[width=0.95\linewidth]{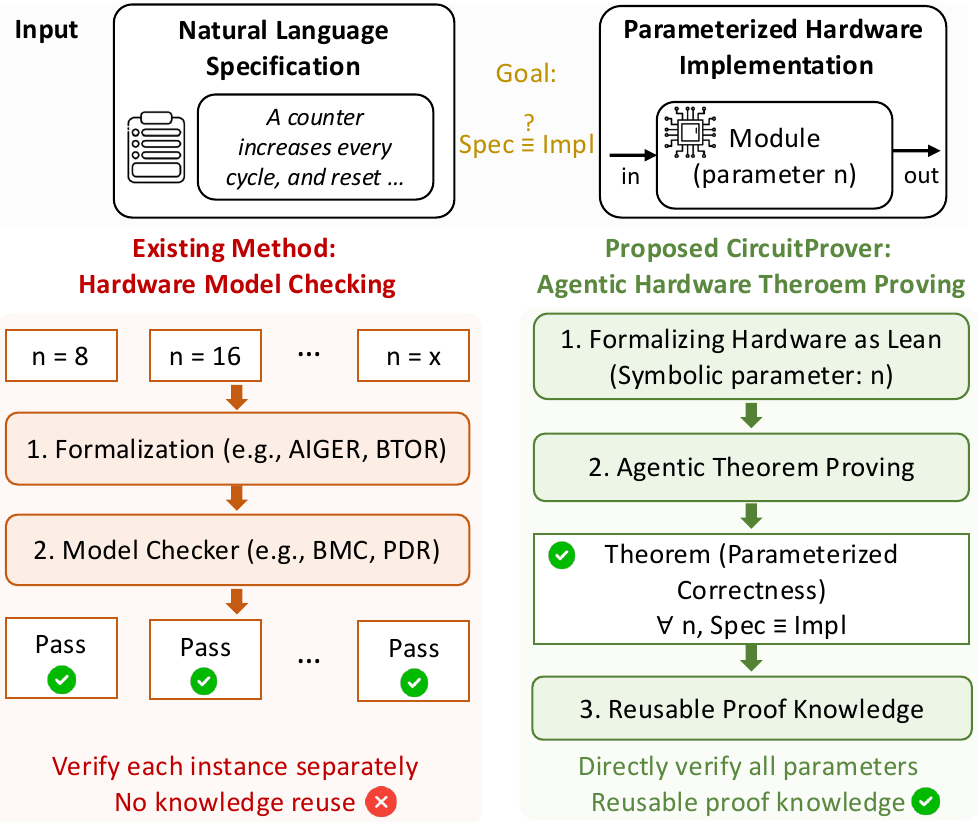}
  \caption{Comparison between existing model checking and our \toolname{} for parameterized hardware verification. \toolname{} proves parameterized correctness in Lean~4 and accumulates reusable proof knowledge, while traditional model checking verifies instantiated designs independently.}
  \label{fig:overview}
\end{figure}

\textbf{Existing verification tools provide automation but lack reusable knowledge.}
Model checking~\cite{clarke2009model, een2011efficient, biere2009bounded, mann2021pono} has become the dominant industrial formal verification methodology. As shown in~\Cref{fig:overview}, it translates hardware designs into solver-friendly transition-system representations, such as AIGER~\cite{brummayer2007aiger}, BTOR~\cite{niemetz2018btor2}, or SMT-LIB~\cite{cok2011smt}, and automatically proves properties or generates counterexamples. This makes model checking highly effective for RTL property checking and bug finding. However, the reasoning behind correctness decisions is largely hidden inside the verification engine and exposed mainly through final pass/fail results or counterexamples. Consequently, verification knowledge is rarely captured as explicit and reusable artifacts, requiring similar reasoning to be reconstructed across related verification tasks. 


\textbf{Interactive theorem proving provides reusable hardware knowledge but lacks automation.}
Beyond automated model checking, interactive theorem proving explicitly represents verification reasoning as machine-checked formal artifacts. 
Existing frameworks such as Kami~\cite{choi2017kami} and Coquet~\cite{braibant2011coquet} in Coq/Rocq, together with ACL2-~\cite{kaufmann1997industrial} and HOL-based~\cite{gordon1988hol} systems, have demonstrated this approach for arithmetic circuits, processors, and parameterized hardware designs. 
These artifacts provide strong correctness guarantees and can be composed and reused across related verification tasks. 
However, current frameworks still rely heavily on experts for hardware formalization, specification development, invariant discovery, and proof construction.

\textbf{What should new hardware verification tools look like in the agent era?}
In the agent era, verification tools should not only check correctness, but also preserve and accumulate explicit proof artifacts and reasoning strategies for future agent reasoning. 
We refer to this paradigm as \textbf{proof-accumulating hardware verification}. 
In this setting, successful proofs are no longer treated as isolated and implicit outputs. Instead, proof strategies, lemmas, and invariants are distilled from proof trajectories, linked to their explicit machine-checked formal artifacts, and reused by agents in new verification tasks.

\textbf{CircuitProver: agentic hardware theorem proving with reusable proof libraries.}
To this end, we present \textbf{CircuitProver}, an agent-native hardware verification framework built upon Lean~4. As shown in~\Cref{fig:overview}, CircuitProver combines hardware formalization, agentic theorem proving, and reusable proof knowledge accumulation. Through proof trajectory distillation, the reasoning processes are transformed into reusable symbolic artifacts that guide future hardware agents.
First, CircuitProver automatically formalizes hardware implementations and specifications into a unified Lean representation for machine-checked reasoning. 
Second, the reasoning agent iteratively constructs and repairs formal proofs through interaction with the theorem prover and its feedback. 
Finally, proving traces and verified theorems are distilled into reusable proof libraries, including natural-language proof strategies and their corresponding machine-checked formal artifacts, and stored in a persistent proof library for future verification tasks.

We study parameterized hardware verification as a representative task for this framework, where a single design generator describes a family of hardware instances with different parameters. 
Existing formal verification methods, such as model checking, often verify individual instantiated designs and may repeatedly encounter similar reasoning patterns across different configurations. 
CircuitProver instead constructs general correctness theorems over symbolic design parameters, enabling a single proof to establish correctness for an entire family of hardware instances.


We summarize the contributions of this work as follows:
\begin{itemize}
\item \textbf{Agent-native hardware verification.}
We present CircuitProver, a Lean~4-based framework that integrates hardware auto-formalization, agentic theorem proving, and reusable proof libraries for automated hardware theorem proving.
\item \textbf{Hardware auto-formalization.}
We develop an automatic formalization pipeline that translates hardware implementations and specifications into a unified Lean~4 representation with machine-checkable semantics.
\item \textbf{Agentic hardware theorem proving.}
We propose an interactive theorem-proving workflow where reasoning agents iteratively construct and repair proofs through feedback from the theorem prover.
\item \textbf{Reusable proof library.}
We build a reusable proof library that distills proof traces into natural-language proof strategies and their corresponding formal artifacts, enabling proof reuse across related verification tasks.
\item \textbf{Benchmark and evaluation.}
We introduce a benchmark suite for agentic hardware theorem proving and demonstrate the effectiveness of \toolname{} and the benefits of reusable proof knowledge.
\end{itemize}

\section{Preliminaries}

\begin{figure*}[!t]
  \centering
  \includegraphics[width=0.98\linewidth]{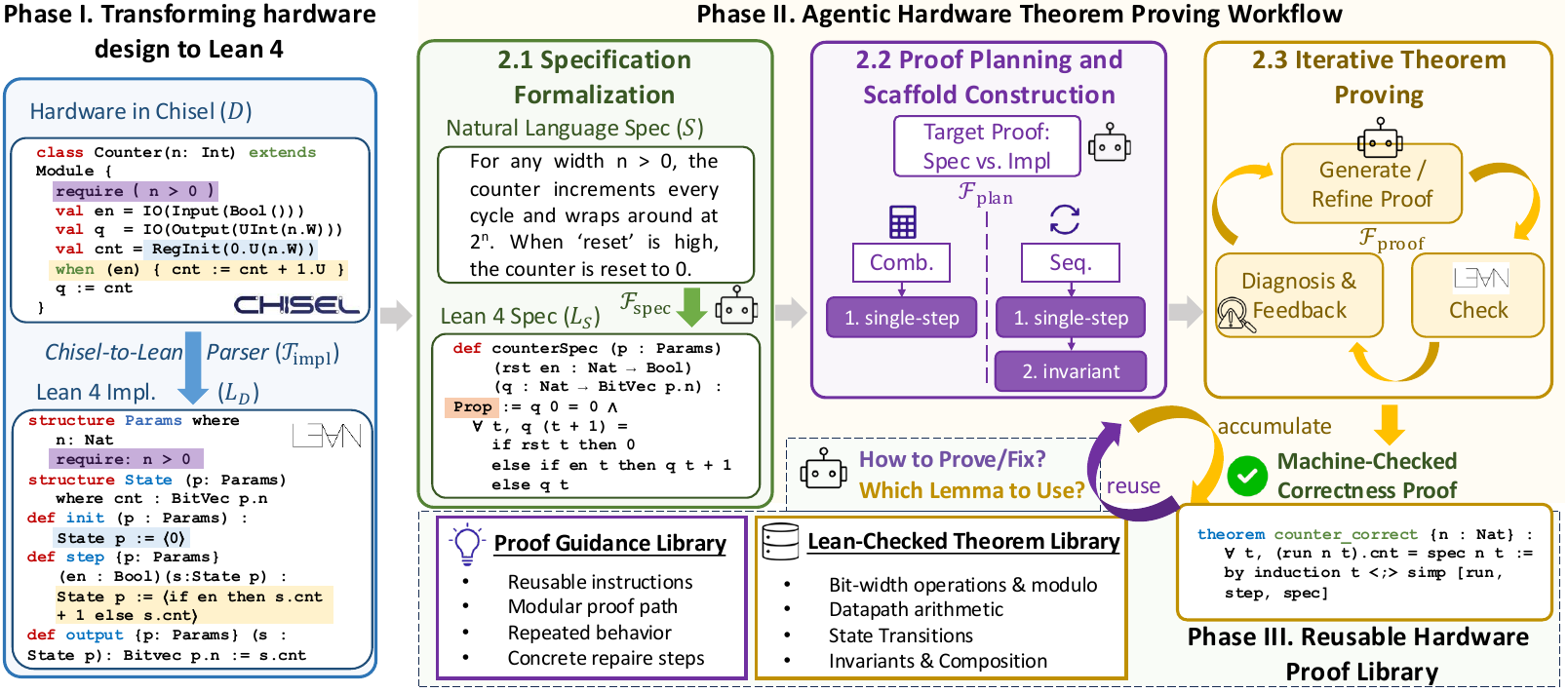}
  \caption{Workflow of \toolname{} for agentic hardware theorem proving. The framework translates hardware designs into Lean~4 models, performs specification formulation, hardware-aware proof planning, and iterative theorem proving, and accumulates reusable proof knowledge.}
  \label{fig:agent}
\end{figure*}

\subsection{Theorem Proving for Hardware Verification}
Hardware theorem proving verifies circuit correctness by encoding hardware implementations and their specifications in an interactive theorem prover and establishing correctness through machine-checked proofs. Early HOL-based verification systems~\cite{gordon1988hol} demonstrated mechanically verified hardware using higher-order logic, while ACL2-based approaches~\cite{kaufmann1997industrial} explored automated theorem proving for hardware reasoning. More recent Coq/Rocq-based frameworks, such as Coquet~\cite{braibant2011coquet} and Kami~\cite{choi2017kami}, enabled embedded hardware modeling and refinement-based verification for parameterized designs. Despite different proof environments, these frameworks share a common workflow of formalizing hardware behavior, specifying correctness properties, and proving corresponding theorems. However, proof development still requires substantial human effort in formalization, invariant discovery, lemma construction, and proof guidance.

\subsection{Lean~4 and Reusable Proof Libraries}
Lean~4 is a modern interactive theorem prover widely used in mathematics, software verification, and AI-assisted theorem proving~\cite{moura2021lean,wu2022autoformalization,azerbayev2023llemma,song2024towards,xin2024deepseek,wang2024lego,li2024hunyuanprover,chen2025automated,kumarappan2025leanagent,ospanov2026apollo}. 
It provides expressive dependent types, programmable proof automation, and a library-based ecosystem; in particular, \texttt{Mathlib}~\cite{mathlib} enables proofs to build upon reusable definitions, lemmas, and automation. 
Recent LLM-based theorem proving frameworks, such as DeepSeek-Prover~\cite{xin2024deepseek} and LeanAgent~\cite{kumarappan2025leanagent}, enable agents to interact with Lean and synthesize proofs for formal statements. 
However, they mainly target mathematical problems and do not address hardware-specific reasoning, which requires circuit-aware proof strategies for fixed-width operations, parameterized structures, and inductive invariants over clocked state transitions. 
Recent work such as CktFormalizer~\cite{xiong2026cktformalizer} explores Lean-based hardware formalization by generating type-checkable circuit representations, while automated proof construction and reusable hardware proof libraries remain largely unexplored.

\section{A Benchmark Suite for Agentic Hardware Theorem Proving}
\label{sec:benchmark}

Before introduing the whole framework, we first introduce a new benchmark suite for evaluating agentic theorem proving on hardware verification.
This benchmark is designed to evaluate whether a theorem-proving workflow can construct checked Lean~4 proofs, and reuse proof knowledge across related verification tasks.
Although the current suite is based on parameterized Chisel designs, its evaluation interface is defined at the Lean level and is therefore independent of a circuit's source language or design style.

\textbf{Benchmark data.}
The main benchmark contains 63 parameterized Chisel verification tasks: 34 arithmetic, 9 control, 9 memory, and 11 miscellaneous. Each task contains the
Chisel implementation, its generated Lean~4 model, a natural-language specification, and a theorem scaffold. The suite covers bit-vector and modular
arithmetic, combinational datapaths, state transitions, and multi-cycle invariants. We additionally isolate four processor-level arithmetic modules---the XiangShan~\cite{wang2025xiangshan} and Rocket~\cite{asanovic2016rocket} multipliers and dividers---to examine behavior on larger designs.

\textbf{Proof tasks.}
Given the hardware design in code format and specification in natural language, the proof agent must construct a complete Lean~4 proof. The target theorem expresses
input-output equivalence for combinational designs or state-transition and invariant properties for sequential designs. A task is solved only when the
final Lean file compiles and contains no \texttt{sorry}, \texttt{admit}, \texttt{axiom}, or \texttt{unsafe}.

\textbf{Evaluation metrics.}
The benchmark evaluates hardware theorem-proving systems along four dimensions.
Proof success is measured by the number and percentage of tasks solved under the acceptance criterion above. 
End-to-end efficiency is measured by wall-clock verification time and, for agentic systems, output tokens and effective proof rounds. 
Proof LoC measures the size and auditability of the resulting machine-checked proof artifact, while Chisel and specification LoC characterize the scale of each task. 
Cost metrics are averaged over the tasks successfully solved by each system, with solved-task counts reported alongside them to make differences in proof success explicit.

\section{Methodology}

\label{sec:transformation}


\subsection{Overview of CircuitProver}

Given a hardware design $D$ and its specification $S$, CircuitProver translates them into Lean~4 representations $L_D$ and $L_S$, and aims to prove a correctness theorem:
\begin{equation}
\vdash_{\mathrm{Lean}} \Phi(L_D,L_S),
\end{equation}
where $\Phi$ specifies that the implementation satisfies the intended behavior. The proof is constructed through auxiliary lemmas, invariants, and intermediate theorems.

As shown in~\Cref{fig:agent}, CircuitProver consists of three components. 
(1) Hardware formalization translates hardware implementations into executable Lean~4 semantic models and formalizes specifications into Lean theorems. 
(2) A reasoning agent iteratively constructs proofs by decomposing proof goals, generating auxiliary lemmas and invariants, and repairing proof scripts based on Lean feedback. 
(3) Successful proof trajectories are distilled into reusable proof knowledge, including reasoning strategies and their corresponding machine-checked artifacts, which are stored in a hardware proof library for future verification tasks.


\subsection{Phase I: Transforming Hardware Design to Lean~4}

As shown in~\Cref{fig:agent}, CircuitProver first translates the hardware implementation into an executable Lean~4 semantic model, providing a fixed and machine-checkable representation for subsequent theorem proving. 
We represent a parameterized hardware design as a transition system:
\begin{equation}
D=(P,I,O,R,T),
\end{equation}
where $P$ denotes design parameters, $I$ and $O$ denote input and output interfaces, $R$ denotes the set of registers, and $T$ defines the state transition behavior.

We develop a deterministic Chisel-to-Lean translator:
\begin{equation}
\mathcal{T}_{\text{impl}}:D\rightarrow L_D,
\end{equation}
where $L_D$ is an executable Lean~4 semantic model of the hardware implementation. 
The generated Lean model preserves the original design parameters, IO interfaces, register states, and state transition behaviors, while representing hardware operations using Lean~4 definitions.

As shown in~\Cref{fig:agent}, the translator extracts design parameters, IO interfaces, registers, and state transitions from the source design and maps them into corresponding Lean~4 definitions. Parameters are represented explicitly, while hardware signals and registers are modeled using width-aware bit-vector types. 
Hardware operations and expressions are translated into executable Lean~4 expressions with consistent width semantics. Intermediate signals are represented using \texttt{let} bindings, conditional logic is translated into \texttt{if} expressions, and register reads and updates are mapped to explicit state accesses and updates in Lean~4.

After translation, we validate the generated models through exhaustive small-bit-width co-simulation: for each configuration, we execute the Chisel design and its Lean~4 counterpart over the complete test space and compare their simulation results. 
The matching results confirm that the translation preserves the behavior of the original design for the evaluated bit-width configurations, providing additional confidence in the generated Lean~4 models before they are used for formal reasoning.

\subsection{Phase II: Agentic Hardware Theorem Proving Workflow}\label{sec:agent}

After the hardware implementation is translated into a Lean~4 model, 
\toolname{} employs a hardware-oriented reasoning agent to construct a machine-checked correctness proof between the implementation and its specification. 
Unlike general theorem proving tasks, hardware verification requires reasoning about parameterized designs, fixed-width bit-vector semantics, and cycle-level state evolution. 
Therefore, \toolname{} develops a hardware-aware proving workflow that integrates circuit analysis, proof planning, and iterative theorem proving with Lean~4 as the trusted proof environment.

As shown in~\Cref{fig:agent}, the workflow consists of three stages: 
(1) specification formulation and circuit analysis, where the agent analyzes the hardware model and constructs the target correctness theorem; 
(2) hardware-aware proof planning and scaffold construction, where the agent identifies appropriate proof structures for different hardware properties and generates auxiliary lemmas and invariants; and 
(3) iterative theorem proving, where the agent interacts with Lean feedback to complete and refine the formal proof.

\textbf{Step 1: Specification formulation.}
Given the transformed Lean hardware model $L_D$ and its natural-language specification $S$, the reasoning agent analyzes the circuit structure and formulates the corresponding Lean~4 specification $L_S$. 
Unlike the deterministic hardware translation in Phase I, this step requires semantic reasoning to interpret the natural-language requirements based on the hardware context. 
The agent analyzes design parameters, parameter constraints, register behavior, reset conditions, output dependencies, and cycle-level behaviors, while preserving parameters symbolically rather than instantiating fixed configurations.

The specification formulation process is defined as:
\begin{equation}
\mathcal{F}_{\mathrm{spec}}:(L_D,S)\rightarrow L_S,
\end{equation}
where $L_S$ consists of formal correctness properties extracted from the natural-language specification. 
Specifically, $L_S$ contains a set of Lean~4 properties:
\begin{equation}
L_S=\{\phi_1,\phi_2,\ldots,\phi_k\},
\end{equation}
where each $\phi_i$ represents a specific correctness requirement to be verified.
We validate the formalized specifications using the same exhaustive small-bit-width co-simulation procedure as in Phase~I.

\textbf{Step 2: Proof planning and scaffold construction.}
Given the formalized hardware model $L_D$ and specification $L_S$, the agent first analyzes the verification task and constructs a hardware-aware proof scaffold. 
Instead of directly searching for tactics on the complete theorem, the agent decomposes the correctness proof into structured intermediate obligations, including auxiliary lemmas and invariants.

The proof planning process is formulated as:
\begin{equation}
\mathcal{F}_{\mathrm{plan}}:(L_D,L_S)\rightarrow \mathcal{P},
\end{equation}
where $\mathcal{P}$ denotes the generated proof scaffold. 
The scaffold construction depends on the hardware behavior:

For \textit{combinational designs}, which do not involve state evolution across cycles, the agent focuses on proving the correctness of the underlying computation within one cycle. 
The generated proof scaffold contains auxiliary lemmas for arithmetic operations, bit-vector manipulations, Boolean logic, and symbolic parameter constraints required to establish functional correctness.

For \textit{sequential designs}, the agent focuses on proving temporal correctness by reasoning about state evolution over time. 
Instead of directly proving the complete execution behavior, the agent first constructs local one-step transition proofs and then discovers the inductive invariants to connect step-wise correctness with global behavioral correctness.

This hardware-aware planning provides structured guidance for subsequent proof construction instead of relying on direct low-level tactic search.

\textbf{Step 3: Iterative theorem proving.}
Given the hardware model $L_D$, specification $L_S$, and generated proof scaffold $\mathcal{P}$, the agent iteratively constructs the final proof through interaction with Lean~4. 
The proof construction process is formulated as:
\begin{equation}
\mathcal{F}_{\mathrm{proof}}:(L_D,L_S,\mathcal{P})\rightarrow \Pi,
\end{equation}
where $\Pi$ denotes the machine-checked proof accepted by the Lean kernel.

In each iteration, the agent updates the proof script, invokes Lean, and analyzes the returned proof states and error messages. 
All generated proof steps are checked by the Lean kernel, ensuring that only formally verified proofs are accepted. 
The agent uses Lean feedback together with the hardware context to refine its reasoning. 
When a proof attempt fails, the agent may revise the proof plan, introduce additional lemmas, strengthen invariants, or adjust assumptions. 
This iterative process continues until all proof obligations are successfully discharged.

\subsection{Phase III: Reusable Hardware Proof Library}
\label{sec:library}

\begin{figure}
    \centering
    \includegraphics[width=\linewidth]{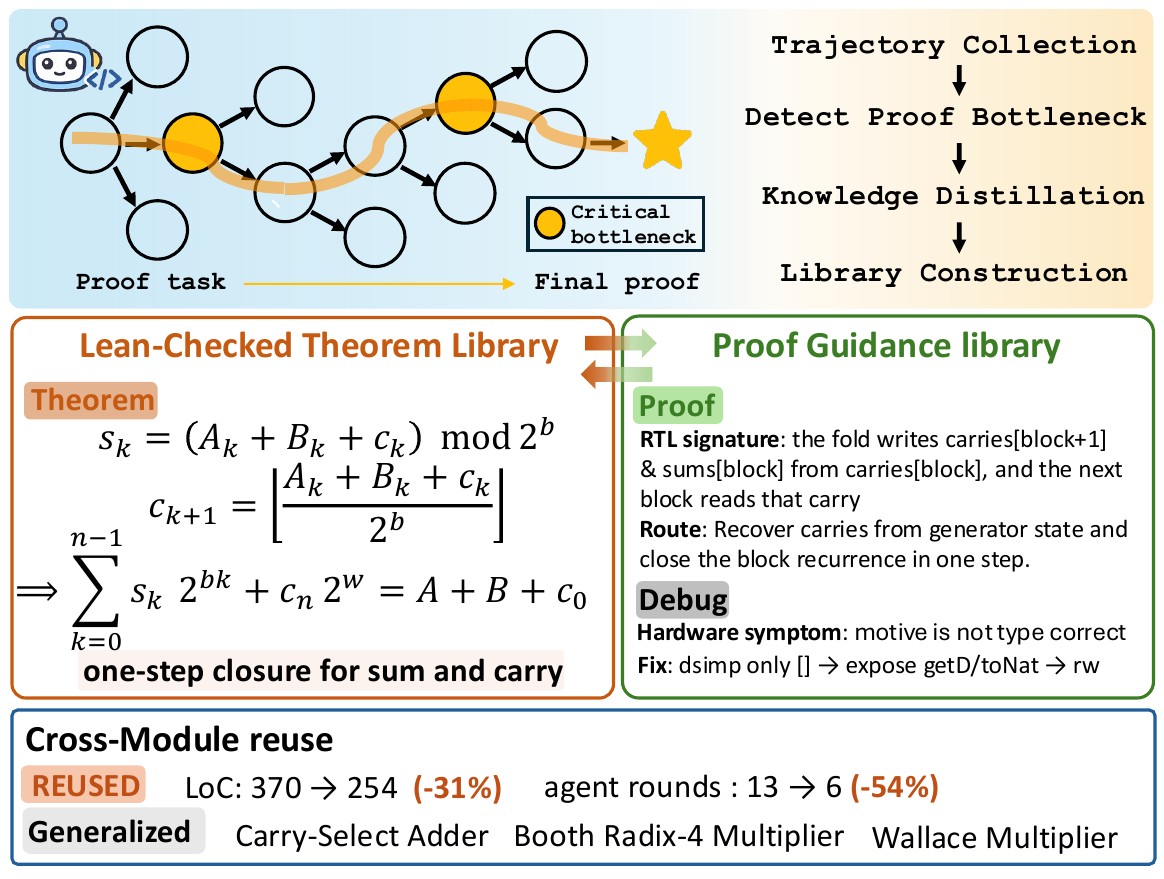}
    \caption{Proof knowledge accumulation in \toolname{}. Proof trajectories are distilled into proof guidance and reusable theorem libraries to accelerate future hardware verification tasks.}
    \label{fig:library}
\end{figure}

As shown in Phase III of~\Cref{fig:agent}, \toolname{} accumulates proof trajectories into two complementary libraries:

\textbf{Proof Guidance Library}. It preserves the knowledge that would otherwise disappear once Lean accepts the final proof, then transforms a successful interaction trajectory into a reusable proof plan that explains how the proof is decomposed, why particular lemmas are selected, and how the plan is adapted in response to Lean feedback.

\textbf{Lean-Checked Theorem Library}. In contrast to the first library, this one preserves the results established along the trajectory. 
Before a theorem is added to the library, \toolname{} decouples it from its originating module by removing module-specific definitions, signal names, and incidental assumptions, and reformulating its reusable core over symbolic parameters and generic interfaces. The resulting theorem is therefore not tied to the module from which
it was extracted and can be instantiated in other modules that exhibit the same underlying hardware semantics or structural pattern.\looseness=-1

\textbf{A concrete example}. \Cref{fig:library} illustrates this two-library organization with concrete entries distilled from the proof of a carry-skip adder.
The \textit{Lean-Checked Theorem Library} stores a module-independent closure theorem that lifts the block-local sum-and-carry relation to the full-width correctness result. 
The corresponding \textit{Proof Guidance Library} entry records when and how to reuse this theorem: it identifies an RTL generator fold in which each
iteration produces a carry consumed by the next iteration, and instructs the agent to recover the carry sequence and apply the closure theorem instead of unfolding the generator step by step. The entry also abstracts the successful fix into a reusable repair principle
for representation-level proof failures: when hidden implementation definitions
prevent a rewrite from matching the proof goal, the agent exposes only the
relevant representation conversions before applying a targeted rewrite.

Together, the two libraries provide both a reusable route for constructing a proof and module-independent, machine-checked facts for completing it, allowing verified knowledge to transfer across hardware designs.

\section{Case Study: Proving Carry-Skip Adder}
\label{sec:case-study}

\begin{figure}
    \centering
    \includegraphics[width=0.95\linewidth]{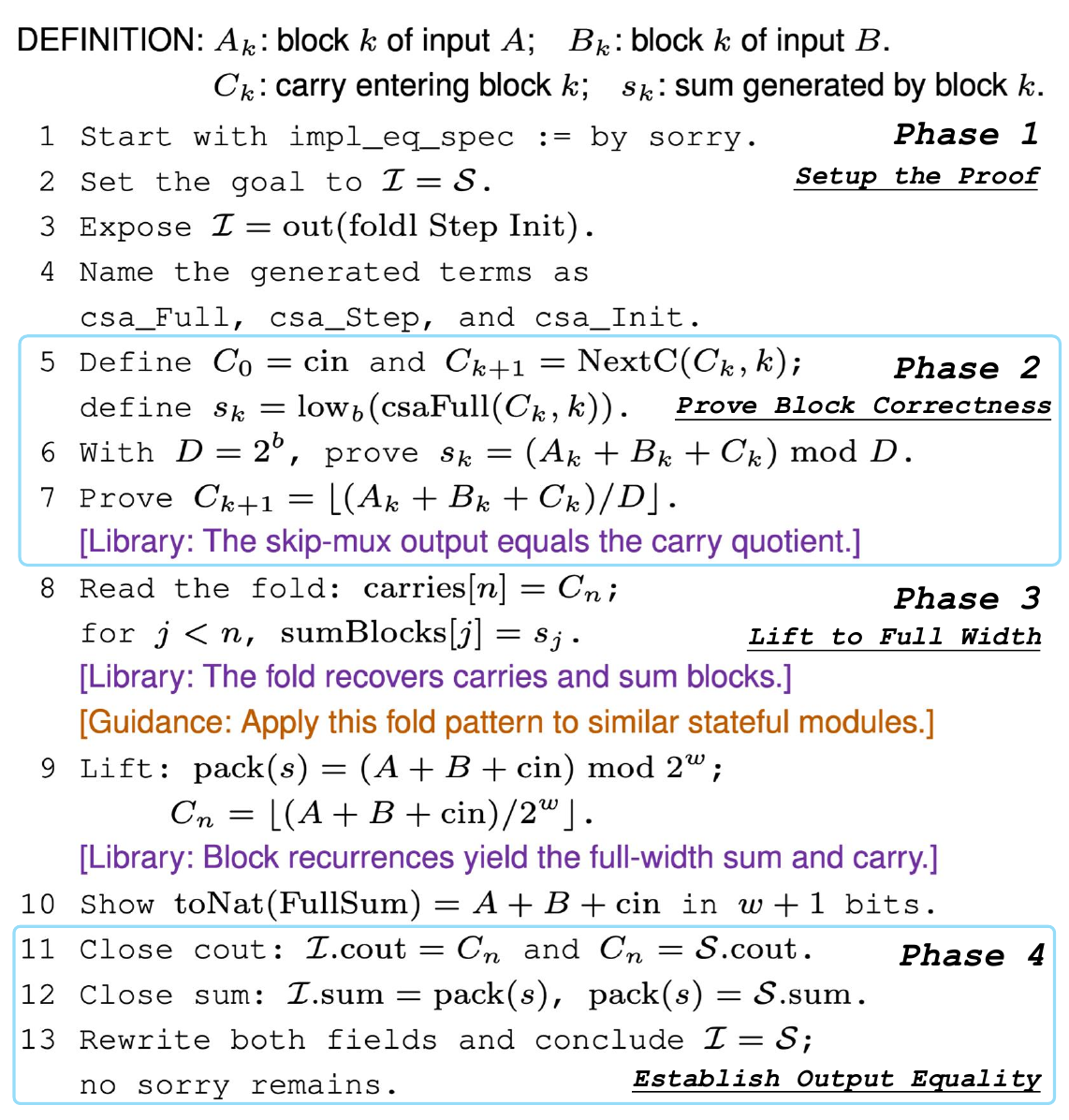}
    \caption{Detailed proof strategy for the carry-skip adder}
    \label{fig:case-study}
\end{figure}

This section uses the carry-skip adder module proof as a case study. This module divides two $w$-bit operands into $n=w/b$ blocks of $b$ bits. Each block computes a widened sum from its two operand slices and the incoming carry. A block propagates the incoming carry when the bitwise XOR of its two operand slices is all ones. The proof path is detailed in~\Cref{fig:case-study}.

\subsection{Four-Phase Proof Strategy}

\textbf{Phase 1: Setup the proof.} 
The target theorem states that the circuit returns the low $w$ bits and bit $w$ of $A+B+\mathrm{cin}$. The generated Lean
model processes blocks from low to high with \texttt{List.foldl}, passing the carry and storing each block sum. 
The proof agent introduces local names for the generated block operation, fold step, and initial state to simplify the proof state.\looseness=-1

\textbf{Phase 2: Prove block correctness.}
Define $C_k$ as the carry entering
block $k$ and $s_k$ as its low $b$-bit sum. For $D=2^b$, both paths of the skip
multiplexer satisfy
$s_k=(A_k+B_k+C_k)\bmod D$ and
$C_{k+1}=\lfloor(A_k+B_k+C_k)/D\rfloor$. The non-skip path obtains the quotient
from the widened addition; on the skip path, the all-ones propagate condition
gives $A_k+B_k=D-1$, so forwarding $C_k$ yields the same quotient.

\textbf{Phase 3: Full-width lifting.}
The proof agent applies a fold read-back lemma to show that the final state
stores $C_n$ and every $s_k$. It then applies the radix closure lemma to compose
the block recurrences into the packed low $w$ bits and the final carry of
$A+B+\mathrm{cin}$. This step lifts the local block equations to the arithmetic
value represented by the widened specification sum.\looseness=-1

\textbf{Phase 4: Establish output equality.} The carry equation identifies
$\mathcal{I}.\mathrm{cout}$ with $\mathcal{S}.\mathrm{cout}$, and a
reverse-Horner lemma rewrites the generated packing fold to identify the two
sum fields. Record extensionality then closes \texttt{impl\_eq\_spec}.

\subsection{Reusable Library Knowledge}

The completed proof yields three reusable insights. First, when every bit in a
block satisfies $A_i \oplus B_i=1$, the two operand slices sum
to $2^b-1$. The outgoing carry therefore equals the incoming carry, which
justifies the carry-skip path. Second, the fold accumulator records the carries
$C_0,\ldots,C_n$ and block sums $s$. Third, the radix-$2^b$
equations combine these block-level facts into the full-width sum and carry.
The library stores these insights as a reusable proof pattern.
For another blocked adder, the agent only needs to re-establish the one-block semantics; the remaining steps can be reused.

\begin{table*}[t]
\centering
\caption{Proof-generation comparison on the full benchmark and processor-level arithmetic modules. Cost metrics are averaged over successfully solved tasks; time is reported in seconds and output tokens in thousands.}
\label{tab:main-baselines}
\small
\renewcommand{\arraystretch}{1.04}
\setlength{\tabcolsep}{1.0pt}
\begin{tabular*}{\textwidth}{@{\extracolsep{\fill}}l*{15}{r}@{}}
\toprule
& \multicolumn{2}{c}{Design LoC} & \multicolumn{3}{c}{Chicala/Stainless$^{\ddag}$} & \multicolumn{5}{c}{Vanilla agent} & \multicolumn{5}{c}{\cellcolor{circuitprover} \textbf{\toolname{} (Ours)}} \\
\cmidrule(lr){2-3}\cmidrule(lr){4-6}\cmidrule(lr){7-11}\cmidrule(l){12-16}
Category (N) & Chisel & Spec & Solved & Time & LoC & Solved & Time & LoC & Tok. & Rnd. & Solved & Time & LoC & Tok. & Rnd. \\
\midrule
\multicolumn{16}{@{}l}{\textit{(a) Full benchmark; four top-level categories}} \\
Arithmetic (34)    & 50.2 & 32.6 & 15/34& 37.2 & 30.6  & 31/34 & 894.8  & 217.7 & 70.3  & 13.0 & 34/34 & 654.7 & 184.1 & 97.8 & 5.7 \\
Control (9)        & 45.0 & 63.3 & 7/9  & 22.9 & 6.4   & 8/9   & 522.9  & 63.9  & 53.2  & 8.9  & 9/9   & 410.7 & 28.8  & 22.8 & 4.3 \\
Memory (9)         & 36.2 & 49.4 & 3/9  & 58.6 & 8.3   & 8/9   & 343.3  & 15.9  & 95.0  & 6.9  & 9/9   & 302.2 & 12.6  & 48.7 & 2.9 \\
Miscellaneous (11) & 27.3 & 27.8 & 6/11 & 16.8 & 13.2  & 11/11 & 63.8   & 8.5   & 30.6  & 3.1  & 11/11 & 53.7  & 7.8   & 10.1 & 2.4 \\
\addlinespace[2pt]
\textbf{Summary (63)} & \textbf{43.4} & \textbf{38.6} & \textbf{31/63} & \textbf{32.1} & \textbf{21.3} & \textbf{58/63} & \textbf{658.9} & \textbf{129.8} & \textbf{66.0} & \textbf{9.2} & \textbf{63/63} & \textbf{506.2} & \textbf{108.6} & \textbf{61.7} & \textbf{4.6} \\
\midrule
\multicolumn{16}{@{}l}{\textit{(b) Processor-level arithmetic modules}} \\
\texttt{xiangshan\_mul} & 377 & 146 & 1/1 & --\runtimemark  & 1675 & 1/1 & 1981.3  & 448 & 278.8 & 9 & 1/1 & 1368.0 & 359 & 231.9 & 13 \\
\texttt{xiangshan\_div} & 73  & 110 & 1/1 & --\runtimemark  & 484 & 1/1 & 2341.1 & 695 & 310.3 & 40 & 1/1 & 251.6  & 456  & 236.5   & 16   \\
\texttt{rocket\_mul}    & 120 & 46 & 1/1 & --\runtimemark     & 507 & 1/1 & 5264.1& 616 &  1334.9 & 54 & 1/1 & 2968.0 & 315 & 252.5 & 45  \\
\texttt{rocket\_div}    & 129 & 63  & 1/1 & --\runtimemark      & 451 & 1/1 & 6685.3 & 1126 & 452.6 & 223 & 1/1 & 2436.7 & 185 & 327.3 & 95 \\
\bottomrule
\end{tabular*}
\begin{itemize}
\item[$\ddag$] Many designs still fail during Chicala compilation.
\item[\runtimemark] Requires substantial manual effort; the exact time cannot be reliably quantified.
\end{itemize}
\end{table*}

\section{Experiment}
\label{sec:experiment}

\subsection{Experiment Setup}
We evaluate \toolname{} on the proposed benchmark suite through three studies: comparisons with Chicala~\cite{feng2024formally} and a matched vanilla agent, component ablations, and language-model scaling.
Our translator currently supports a subset of Chisel that covers the common language constructs. 
The frontend adapts the typed-AST translation pipeline introduced by Chicala and builds a new parsing and translation pipeline around this representation. \toolname{} is performed with Lean~4 v4.29.0.
Our current prototype runs the agent through the Claude Code command-line interface (CLI), but the workflow itself is agent-CLI agnostic: the task specification and repair procedure are defined independently of the agent, and proof knowledge is provided through a reusable theorem library.
All experiments are run on a server with Ubuntu 20.04.4 LTS, dual Intel Xeon Platinum 8375C processors, and 256GB RAM.

\subsection{Main Results: Proof-Generation Comparison }
Table~\ref{tab:main-baselines} compares the proof-generation efficiency and cost of \toolname{} with Chicala/Stainless and the vanilla agent. 
The vanilla agent is a matched baseline that uses the same language model, generated Lean model, specification, and theorem scaffold as \toolname{}, but does not have access to its reusable hardware theorem library or hardware-aware proof guidance.

\textbf{Overall module proving results.}~\Cref{tab:main-baselines}(a) presents the main benchmark results across the four circuit categories.
The vanilla agent solves all miscellaneous tasks but misses cases in arithmetic, control, and memory, indicating limitations in general proof search. Chicala/Stainless is limited earlier by front-end robustness: despite our repair of more than 20 implementation issues, many designs still fail during compilation and therefore never reach Stainless verification. 

\textbf{Proof-generation efficiency results.} Beyond improving proof success, \toolname{} turns proof generation from repeated
trial-and-error search into a hardware-guided process that can reuse previously
established lemmas and proof strategies. Compared with the vanilla agent, 
it reduces the average number of proof rounds by 50.0\%, verification time by 23.2\%, and proof LoC by 16.3\%. 
These reductions indicate that the agent reaches checked proofs with fewer repair cycles and less
redundant proof construction. Output tokens decrease more modestly, from 66.0K
to 61.7K, and increase for arithmetic because \toolname{} solves three
additional difficult arithmetic tasks; the token averages therefore cover
different sets of solved tasks. Overall, the results support the central premise
of proof-accumulating verification: reusable hardware knowledge improves both
proof success and proof-search efficiency.\looseness=-1

\textbf{Processor-level module proving results.} Table~\ref{tab:main-baselines}(b) further suggests that these benefits grow with design complexity. With identical coverage on the four processor-level modules, \toolname{} reduces aggregate verification time, proof LoC, output tokens, and proof rounds by 56.8\%, 54.4\%, 55.9\%, and 48.2\%, respectively, relative to the vanilla agent, and reduces aggregate proof LoC by 57.8\% relative to Chicala/Stainless. 
The pronounced reductions on these larger designs indicate that hardware-aware proof decomposition and reusable proof knowledge become increasingly valuable as it becomes more expensive to rediscover invariants and auxiliary lemmas and to repair proofs.

\subsection{Ablation Study}

\begin{figure}[t]
\centering
\includegraphics[width=\linewidth]{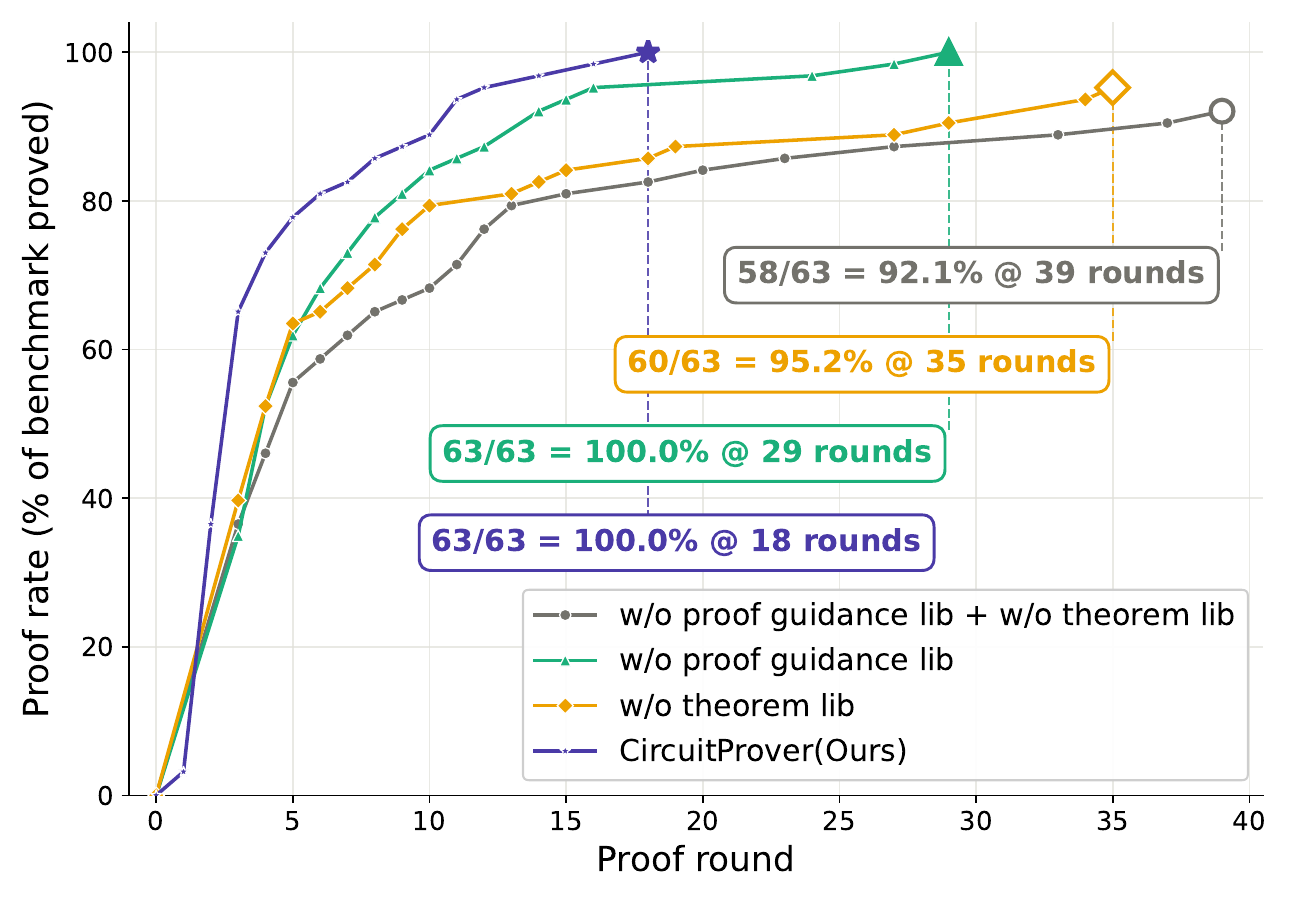}
\caption{Ablation study of \toolname{}, evaluating the impact of proof guidance and Lean-checked theorem libraries.}
\label{fig:ablation}
\end{figure}

To isolate the contributions of the reusable theorem library and hardware-aware
proof guidance, we conduct a controlled $2\times2$ ablation over the same 63
tasks. All four configurations use the same language model, Lean model,
specification, and theorem scaffold. Figure~\ref{fig:ablation} reports the
cumulative number of solved tasks under a 40-round budget, with each round
corresponding to one Lean compilation.

The $2\times2$ ablation exposes distinct roles for the two components. Holding
guidance fixed, adding the library raises within-budget coverage from 60 to 63
tasks; without guidance, the gain increases from 58 to 63. The library
therefore improves within-budget reachability rather than merely shifting the
same solutions earlier. Holding the library fixed instead, guidance advances
the number of solved tasks at round 10 from 53 to 56 and reduces the rounds
required for full coverage from 29 to 18. The same acceleration appears without
the library, where guidance increases the round-10 result from 43 to 50 tasks,
although neither configuration reaches full coverage. These paired contrasts
show that reusable lemmas close difficult cases, while hardware-aware guidance
shortens the search needed to exploit them; their combination achieves both
full coverage and the fastest convergence.

\subsection{Scaling with LLM Capability}

To assess sensitivity to language-model capability, we evaluate \toolname{}
with four Claude models under the same benchmark and workflow.
Table~\ref{tab:model-comparison} shows that proof coverage rises from 31/63 with
Haiku 4.5 to 47/63 with Sonnet 4.6, 60/63 with Sonnet 5, and 63/63 with Opus
4.8. Opus is the only model to solve every task and also has the lowest average
time, proof LoC, token usage, and proof rounds. Although these cost averages
cover different solved-task subsets, the results show that stronger models
improve proof coverage and, at the high end, proof-search efficiency.

\begin{table}[t]
\centering
\caption{Comparison of \toolname{} with different Claude models on the benchmark.}
\label{tab:model-comparison}
\small
\renewcommand{\arraystretch}{1.05}
\setlength{\tabcolsep}{3.0pt}
\begin{tabular}{@{}lrrrrr@{}}
\toprule
Model & Solved $\uparrow$ & Time (s) $\downarrow$ & LoC $\downarrow$ & Tok. (K) $\downarrow$ & Rnd. $\downarrow$ \\
\midrule
Haiku 4.5  & 31/63 & 845.2 & 174.3 & 301.5 & 17.5 \\
Sonnet 4.6 & 47/63 & 834.5 & 139.0 & 480.4 & 11.5 \\
Sonnet 5 & 60/63 & 642.1 & 113.4 & 88.9 & 11.7 \\
Opus 4.8   & 63/63 & 506.2 & 108.6 & 61.7 & 4.6 \\
\bottomrule
\end{tabular}
\end{table}

\section{Conclusion}

This paper presented \toolname{}, an agentic Lean~4 framework for
proof-accumulating verification of parameterized Chisel designs. Experiments
demonstrate that \toolname{} is effective and efficient across diverse hardware
verification tasks. Ablations show that reusable lemmas help solve difficult
cases, while hardware-aware guidance accelerates search. These results show
that proof accumulation benefits hardware theorem proving. 
Future work will extend support to broader RTL constructs, richer properties, and industrial designs.\looseness=-1


\bibliography{aaai2027}




\end{document}